\documentstyle[aas2pp4,psfig]{article}

\newcommand\mdot   {\hbox {${\dot M}$}}

\newcommand\pp     {$\pm$}

\newcommand\micros  {$\mu$s}

\righthead{KiloHertz Quasi-Periodic Oscillations in 4U 1735$-$44}
\slugcomment{Submitted to ApJ Letters, November 1997}

\begin{document}

\title{Discovery of KiloHertz Quasi-Periodic Oscillations in 4U 1735$-$44}

\author{Rudy Wijnands\altaffilmark{1},
        Michiel van der Klis\altaffilmark{1},
	Mariano M\'endez\altaffilmark{1,2},
	Jan van Paradijs\altaffilmark{1,3},
	Walter H. G. Lewin\altaffilmark{4},
	Frederick K. Lamb\altaffilmark{5},
        Brian Vaughan\altaffilmark{6},
	Erik Kuulkers\altaffilmark{7}
        }

\altaffiltext{1}{Astronomical Institute ``Anton Pannekoek'',
University of Amsterdam, and Center for High Energy Astrophysics,
Kruislaan 403, NL-1098 SJ Amsterdam, The Netherlands;
rudy@astro.uva.nl, michiel@astro.uva.nl, mariano@astro.uva.nl,
jvp@astro.uva.nl}

\altaffiltext{2}{Facultad de Ciencias Astron\'omicas y
Geof\'{\i}sicas, Universidad Nacional de La Plata, Paseo del
Bosque S/N, 1900 La Plata, Argentina}

\altaffiltext{3}{Departments of Physics,
University of Alabama at Huntsville, Huntsville, AL 35899}

\altaffiltext{4}{Department of Physics and Center for Space Research,
Massachusetts Institute of Technology, Cambridge, MA 02139;
lewin@space.mit.edu}

\altaffiltext{5}{Departments of Physics and Astronomy,
University of Illinois at Urbana-Champaign, Urbana, IL 61801;
f-lamb@uiuc.edu}

\altaffiltext{6}{Space Radiation Laboratory, California
Institute of Technology, 220-47 Downs, Pasadena, CA 91125;
brian@thor.srl.caltech.edu}

\altaffiltext{7}{Astrophysics, University of Oxford,
Nuclear and Astrophysics Laboratory, Keble Road, Oxford OX1 3RH,
United Kingdom; e.kuulkers1@physics.oxford.ac.uk}

\begin{abstract}

We discovered a single kHz quasi-periodic oscillation (QPO) near 1150
Hz in the Rossi X-ray Timing Explorer X-ray light curve of the low
mass X-ray binary and atoll source 4U\,1735$-$44. The rms amplitude of
this peak was 2--3 \%, and the FWHM 6--40 Hz. There are indications
that the kHz QPO frequency decreased from 1160 Hz to 1145 Hz when the
count rate increased, which would be quite different from what is
observed in other atoll sources for which kHz QPOs have been
discovered. In the X-ray color-color diagram and hardness-intensity
diagram the source traced out the curved branch (the so-called banana
branch) which has been found by previous instruments. The kHz QPO was
only detected when the source was at the lowest count rates during our
observations, i.e. on the lower part of the banana branch. When 4U
1735$-$44 was at higher count rates, i.e. on the upper part of the
banana branch and at higher inferred mass accretion rate with respect
to that on the lower part of the banana branch, the QPO was not
detected.

Besides the kHz QPO we discovered a low frequency QPO with a frequency
near 67 Hz, together with a complex broad peaked noise component below
30 Hz. This 67 Hz QPO may be related to the magnetospheric
beat-frequency QPO, which is observed on the horizontal branch of Z
sources. This idea is supported by the (peaked) noise found in both 4U
1735$-$44 and Z sources at frequencies just below the QPO frequency.

\end{abstract}


\section{Introduction \label{intro}}

The low-mass X-ray binaries (LMXBs) can be divided into two
subclasses, the atoll sources and the Z sources, after the track they
trace out in the X-ray color-color diagram (CD) (Hasinger \& van der
Klis 1989).  Atoll sources trace out an atoll like shape in the CD
(see e.g. Fig. 6.9 of van der Klis 1995), with a curved branch called
the banana branch (divided in the upper part and lower part of the
banana branch according to the position in the CD), and one or more
patches corresponding to the island state, in which the colors do not
change much on time scales of a day.  The exact morphology of the
island state track, if any, is unknown as it apparently takes a long
time to trace it out.  Atoll sources have lower mass accretion rates
(\mdot) than Z sources (e.g. Hasinger \& van der Klis 1989).  In atoll
sources \mdot\, varies considerably between sources, and even within a
single source. It is thought that \mdot\, is lowest when the sources
are in the island state, increases on the banana branch and highest on
the upper part of the banana branch. The properties of the rapid X-ray
variations that are observed in these sources are correlated with
their position on the atoll. When the sources are in the island state
strong band-limited noise is detected (called high frequency noise or
HFN, with cut-off frequencies between 10 and 30 Hz). This noise is
much weaker when atoll sources are on the lower part of the banana
branch and it is undetectable when they are on the upper part of the
banana branch.
 
Quasi-periodic oscillations (QPOs) at kHz frequencies have been
observed in a large number of LMXBs (see van der Klis 1997 for a
recent review).  The kHz QPOs in atoll sources are usually found when
the sources are in the island state and, rarely, when they are on the
lower part of the banana branch. When they are on the upper part of
the banana branch so far no kHz QPOs have been detected, with
stringent upper limits (Wijnands et al. 1997; Smale, Zhang, \& White
1997). Usually, two simultaneous kHz QPOs are detected, whose
frequencies increase with \mdot.  Sometimes, a single kHz QPO is
detected (4U 1636$-$53: Zhang et al. 1996; Wijnands et al. 1997; Aql
X-1: Zhang et al. 1997).  In this letter, we report the discovery of a
single kHz QPO near 1150 Hz in the atoll source 4U 1735$-$44.  A
preliminary announcement of this discovery was made by Wijnands et
al. (1996).

\section{Observations and Analysis  \label{observations}}

We observed 4U 1735$-$44 on Aug 1, Sep 1, 4, and 28, and Nov 29, 1996
using the proportional counter array (PCA) onboard the Rossi X-ray
Timing Explorer (RXTE; Bradt, Rothschild \& Swank 1993), and obtained
a total of 27 ksec of data. During part of the Sep 28 observation only
4 of the 5 PCA detectors were on. During all observations data were
collected in 129 photon energy channels (effective energy range 2--60
keV) with a time resolution of 16 seconds. Simultaneously, data were
collected during the Aug 1, Sep 28, and Nov 29 observations in one
broad energy channel (effective range: 2--18.2 keV) with a time
resolution of 122 \micros.  During the Sep 1 and 4 observations data
were collected in three broad energy bands with a time resolution of
122 \micros\, (total effective energy band: 2--17.8 keV).

For the analysis of the X-ray spectral variations we used the 16s data
of the four detectors which were always on.  In constructing the CD we
used for the soft color the count rate ratio between 3.9--6.0 keV and
2.0--3.9 keV, and for the hard color the ratio between 8.6--18.9 keV
and 6.0--8.6 keV. For the hardness-intensity diagram (HID) we used as
intensity the count rates in the energy band 2.0--18.9 keV, and for
the hardness the same as we used for the hard color in the CD.  For
the analysis of the rapid time variability we made power density
spectra of all the available 2--18.2 keV and 2--17.8 keV data, and
combined these.  For determining the properties of the kHz QPOs we
fitted the 128--2048 Hz power spectra with a function consisting of a
constant and one or two Lorentzian peaks. For measuring the high
frequency noise and the low frequency QPO we fitted the 0.1-512 Hz
power spectra with a constant, an exponentially cut-off power law, and
one or two Lorentzian peaks. The errors on the fit parameters were
determined using $\Delta\chi^2 = 1.0$; the upper limits were
determined using $\Delta\chi^2 = 2.71$, corresponding to a 95 \%
confidence level. Upper limits on the kHz QPOs were determined using a
fixed FWHM of 50 Hz.

\section{Results \label{results}}

\subsection{Atoll source state\label{state}}

The CD and HID for all data combined are shown in
Fig.\,\ref{CD_HID}. The source traced out a clear banana branch in
both diagrams. There is more scatter on the upper part of the banana
branch in the HID than in the CD, which is due to the fact that during
the Sep 1 and 4 observations the count rate when the source was on the
upper part of the banana branch was slightly higher than during the
other observations. From the CD and the strength of the HFN (see
Section\,\ref{lowfreq}) it is clear that 4U 1735$-$44 was never in the
island state during our observations.

\subsection{KiloHertz QPO \label{khz}}

\begin{figure}[t]
\begin{center}
\begin{tabular}{c}
\psfig{figure=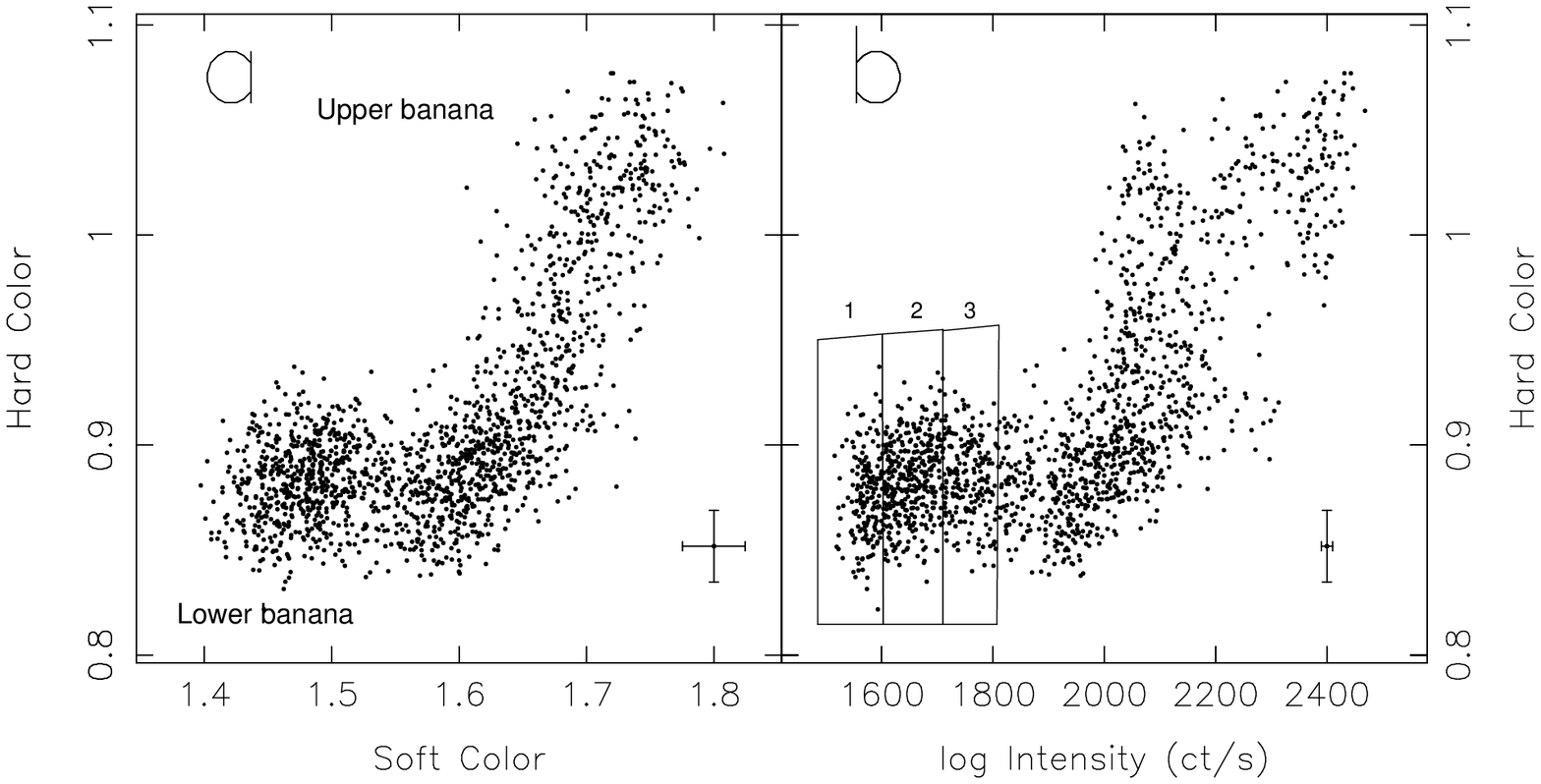,width=18cm}
\end{tabular}
\caption{Color-color diagram ({\it a}) and
hardness-intensity diagram ({\it b}) of 4U 1735$-$44. The soft color
is the count rate ratio between 3.9--6.0 keV and 2.0--3.9 keV; the
hard color is the count rate ratio between 8.6--18.9 keV and 6.0--8.6
keV; the intensity is the 4-detector count rate in the photon energy
range 2.0--18.9 keV. The count rates are background subtracted, but
not dead-time corrected (typically 1\% correction). All points are 16
s averages.  In the HID the three boxes are shown which were used to
select power spectra to study the timing properties. Typical error
bars on the colors and intensity are plotted in the lower right corner
of the diagrams.
\label{CD_HID}}
\end{center}
\end{figure}

We selected power spectra according to the position of the source on
the banana branch in the HID (thus effectively by count rate level in
the HID). In the combined data of regions 1, 2, and 3 (see
Fig. \ref{CD_HID}b) we detected a kHz QPO at 1149\pp4 Hz at an
amplitude that differed from zero by 5.6 times the 1$\sigma$
uncertainty (Table \ref{tab1}), quite significant when taking into
account the $\sim$30 trials implicit in our procedure to search the
power spectrum out to 1200 Hz. The QPO was not detected in the rest of
the data. In order to determine the behavior of the kHz QPO with
changing position of the source in the HID we divided the region in
which the kHz QPO was found into three parts (see Fig. \ref{CD_HID}b
and Tab.\,\ref{tab1}). In the data corresponding to regions 1 and 2 we
detected the kHz QPO at 1161\pp1 Hz (3.3$\sigma$) and 1144\pp4 Hz
(4.8$\sigma$), respectively. In the data of region 3 the kHz QPO was
not detected.

\begin{figure}{t}
\begin{center}
\begin{tabular}{c}
\psfig{figure=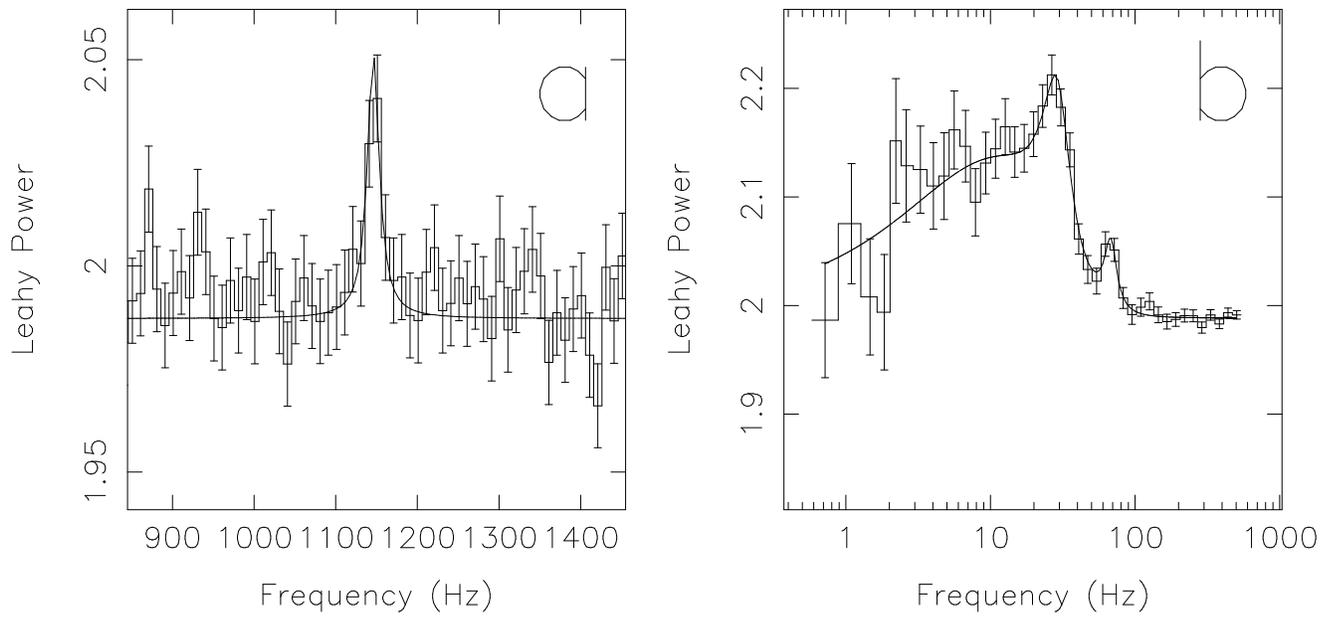,width=18cm}
\end{tabular}
\caption{Typical Leahy normalized power spectra
in the energy range 2--18 keV. {\it a} shows the kHz QPO; {\it b} the
complex high frequency noise and the 67 Hz QPO. Both power spectra
were selected for the time interval 1996 August 1 15:12--16:15 UTC.
\label{powerspectrum}}
\end{center}
\end{figure}

We also made power spectrum selections based on continuous time
intervals (mainly different RXTE data segments). In the first 8000 s
(5040 s of data; 1996 Aug 1 14:17--16:15 UTC) of the observation on
August 1 we detected a kHz QPO at 1149\pp3 Hz (4.5$\sigma$). We
divided this data segment in the intervals 0--2000s (1296 s of data)
and 2000--8000s (3744 s of data) and detected the 1160 Hz QPO in the
first interval (4.0$\sigma$; 1567\pp25 ct/s for 2--18.9 keV) and the
1145 Hz QPO in the second interval (4.8$\sigma$; 1623\pp49 ct/s for
2--18.9 keV; see Fig.\,\ref{powerspectrum}a). The 1145 Hz QPO was also
detected in the time interval 1996 Sep 28 12:50--13:49 UT
(3.8$\sigma$; 3520 s of data).

In both cases, the kHz QPO near 1160 Hz was the narrowest and least
significant.  Applying an $F$-test to the $\chi^2$ of the fits with
and without this QPO, we obtain probabilities of $3.0\times10^{-5}$
and $1.3\times10^{-5}$, for the count rate selection data and the time
selection data respectively, for the hypothesis that the 1160 Hz QPO
was not present in the data. As the effective number of trials
implicit in our search for a peak near 1149 Hz was about 14, the 1160
Hz QPO was detected at high confidence.

We tested the significance of the frequency change between HID regions
1 and 2, and between the 0--2000 s and 2000--8000 s intervals by
fitting simultaneously the two power spectra, and either forcing the
frequency to be the same in both fits or leaving both frequencies
free.  Applying an $F$-test to the $\chi^2$ values of these fits we
obtain probabilities of $2\times10^{-2}$ and $3\times10^{-4}$,
respectively, for the hypothesis that the frequency of the kHz QPOs
did not change in the sense reported. The probability differs
considerably between the two selection methods. Although there are no
trials in this determination this frequency shift needs further
confirmation

It is known from other LMXBs that the kHz QPOs are strongest at the
high photon energies. However, due to the lack of energy resolution of
the high timing data when the kHz QPO was present we could not check
this for 4U 1735$-$44.

In the power spectrum of the combined region 1, 2, and 3 we detected
simultaneously with the 1149 Hz QPO a second peak at 900\pp14 Hz (FWHM
of 68\pp29 Hz; amplitude of 2.6\pp0.4\% rms; 3.4$\sigma$). Applying an
$F$-test to the $\chi^2$ of the fit with and without the 900 Hz QPO we
obtain a probability of $2.2\times10^{-4}$ for the hypothesis that the
900 Hz QPO was not present in the data, indicating that the 900 Hz QPO
was detected at a 3.8$\sigma$ level.  Taking into account the number
of trials ($\sim$8) this QPO is only marginally significant. Shifting
the 1160 Hz QPO to 1140 Hz and averaging all QPO data together
(cf. M\'endez et al. 1997b) did not improve the significance of the
second QPO.  If this second QPO is really present the frequency
separation would be $\sim$250 Hz, not uncommon as a peak difference in
LMXBs.

\subsection{Low-frequency power spectrum \label{lowfreq}}

In the power spectrum below 100 Hz for the time interval 1996 Aug 1
15:12--16:15 UTC a 67 Hz QPO and peaked noise below 30 Hz were clearly
detected (Fig.\,\ref{powerspectrum}b). The peaked noise component had
a complex form. We fitted this noise component with an exponentially
cut-off power law plus a Lorentzian.  Including this Lorentzian in the
fits considerably improved the quality of the fits, and made the
properties of the 67 Hz QPOs much easier to determine.  The rms
amplitude, power law index, and cut-off frequency of the cut-off power
law were 3.7\pp0.7\%, $-$1.1\pp0.5, and 8\pp3 Hz. The rms amplitude,
FWHM, and frequencies of the extra Lorentzian and the 67 Hz QPO were,
5.1\pp0.7\% (3.9 $\sigma$) and 3.0\pp0.3\% (4.9 $\sigma$), 17\pp3 Hz
and 17\pp5 Hz, and 28.7\pp0.9 Hz and 67\pp2 Hz, respectively.  The 67
Hz QPO and the noise component were best detected when 4U 1735$-$44
was on the lower part of the banana branch, at count rates below 1800
ct/s (regions 1, 2, and 3 of Fig.\,\ref{CD_HID}b).  The behavior of
the properties of the 67 Hz QPO with count rate and position of the
source on the banana branch could not be accurately determined.
However, it was clear that the 67 Hz QPO and the peaked noise
component are weaker at somewhat higher count rates and not detected
when the source is on the upper part of the banana branch.

\section{Discussion \label{discussion}}

We detected for the first time a kHz QPO in 4U 1735$-$44. Its
frequency was near 1149 Hz. A second kHz QPO might be present at
$\sim$900 Hz, but this needs confirmation. The kHz QPO was only
detected when the source was at the lowest count rates during our
observations (i.e. on the lower part of the banana branch).  There is
an indication that the kHz QPO frequency decreased with increasing
count rate, which would be quite different from what is observed in
other LMXBs showing kHz QPOs.  Most kHz QPOs increase in frequency
with increasing \mdot. Perhaps the small frequency decrease indicates
that the inner accretion disk reached the innermost stable orbit
during our observations. When the inner disk reaches the innermost
stable orbit it is expected (Kaaret, Ford, \& Chen 1997; Miller, Lamb,
\& Psaltis 1997) that the increase in kHz QPO frequency with \mdot\,
levels off. The frequency of the kHz QPO could then change erratically
(Kaaret et al. 1997), and a small frequency decrease could be
observed.

On the basis of its luminosity and its burst properties (e.g. van
Paradijs et al. 1979; Lewin et al. 1980; van Paradijs et al. 1988) 4U
1735$-$44 is believed to have a higher \mdot\, than the lower
luminosity atoll sources (e.g. 4U 0614$+$4192, 4U 1728$+$28), but a
lower \mdot\, than the more luminous sources (e.g. GX 9$+$9, GX
9$+$1). Thus, 4U 1735$-$44 is believed to have intermediate \mdot\,
for an atoll source. The detection of only a weak kHz QPO, only at low
\mdot, supports this. So far, for the atoll sources with the highest
\mdot\, no kHz QPOs have been found, with upper limits of typically
1--2\% (GX 9$+$9, GX 9$+$1: Wijnands, van der Klis, \& van Paradijs
1997; GX13$+$1: Homan et al. 1997; GX3$+$1: Strohmayer et
al. 1997). For the sources with lower \mdot, kHz QPOs are often
detected with rms amplitudes of 6--11 \% in the total PCA energy range
(see e.g. Strohmayer et al. 1996; Zhang et al. 1996; Berger et
al. 1996; M\'endez et al. 1997a). The weak kHz QPO in 4U 1735$-$44
detected only at the lowest count rates during our observations
(i.e. when the source was on the lower part of the banana branch) fits
naturally in this scheme. We expect that if \mdot\, for 4U 1735$-$44
would drop below our lowest level the kHz QPOs would become stronger
and two peaks would be detected simultaneously.

KHz QPOs in atoll sources are usually found when the sources are at
low inferred \mdot, i.e.  when they are in the island state
(Strohmayer et al. 1996; M\'endez et al. 1997a; Yu et al. 1997) and on
the lower part part of the banana branch (Wijnands et al. 1997, Smale
et al. 1997).  At the lowest inferred \mdot\, in the island state of
4U 0614$+$09 no kHz QPOs were detected.  So far, no kHz QPOs have been
detected when the sources are at higher inferred \mdot, i.e. when they
are on the upper part of the banana branch, with stringent upper
limits (Wijnands et al. 1997; Smale et al.  1997). The kHz QPO in 4U
1735$-$44 is only detected when the source was at the lowest count
rates during our observations (on the lower part of the banana branch)
and undetectable at higher count rates (further up the banana branch,
at higher inferred \mdot). It seems that the kHz QPOs in atoll sources
are present at low \mdot, but possible not at the lowest \mdot\,
(M\'endez et al. 1997a), and disappear when \mdot\, increases.

Besides the kHz QPO we detected a QPO at 67 Hz.  Also, peaked noise
with a complex form below 30 Hz is detected, obviously the well-known
high frequency noise (HFN) in atoll sources (Hasinger \& van der Klis
1989).  The HFN was not detected when the source was on the upper part
of the banana branch (at the highest count rates during our
observations), and the 67 Hz QPO already disappeared when the source
was in the middle part of the banana branch. Similar low-frequency
QPOs were detected in several other atoll sources (e.g. Hasinger \&
van der Klis 1989; Strohmayer et al. 1996; Wijnands \& van der Klis
1997; Homan et al. 1997). It is possible that the 67 Hz QPO detected
in 4U 1735$-$44 is similar to the horizontal branch QPOs (HBOs) in Z
sources.  This is supported by the facts that the frequency of this
QPO is similar to the frequencies of the HBOs in Z sources (although a
fundamental frequency as high as 67 Hz has never been observed for a
HBO in Z sources), and that it is accompanied by a (peaked) noise
component similar to what is found in the Z sources (the so-called
low-frequency noise). Van der Klis (1994) already suggested that the
HFN in atoll sources is due to the same physical process as the LFN in
Z sources.

Recently, Stella \& Vietri (1998) proposed that the low frequency QPOs
observed in atoll sources are due to a precession of the innermost
disk region, dominated by the Lense-Thirring effect. If we assume that
the 900 Hz QPO is real then the frequency difference would be 249
Hz. Following their reasoning and assumptions, and assuming that the
neutron star spin frequency is this frequency difference, we derive a
maximum precession frequency of $\sim$28 Hz. The 67 Hz QPO we
discovered in 4U 1735$-$44 can not easily be explained by this model
without an $I/M$ (with $I$ is the moment of inertia of the neutron
star and $M$ the neutron star mass) which is more than 2 times larger
than allowed for any mass and equation of state. However, the complex
nature of the HFN, and especially the peak near 29 Hz, could be due to
the precession of the inner disk.

\acknowledgments

This work was supported in part by the Netherlands Foundation for
Research in Astronomy (ASTRON) grant 781-76-017 and by NSF grant AST
93-15133. B.V. (NAG 5-3340), F.K.L (NAG 5-2925), J. v. P. (NAG 5-3269,
NAG 5-3271) and W. H. G. L.  acknowledge support from the United
States NASA grants. MM is a fellow of the Consejo Nacional de
Investigaciones Cient\'{\i}ficas y T\'ecnicas de la Rep\'ublica
Argentina.

\clearpage

\clearpage

\clearpage

\begin{deluxetable}{ccccc}
\tablecolumns{5}
\tablewidth{0pt}
\tablecaption{kHz QPO properties$^a$ \label{tab1}}
\tablehead{
Selection $^b$    & Count rate$^c$ &rms$^d$             & FWHM     & Frequency\\
              & (ct/s)    &(\%)                & (Hz)    & (Hz)}
\startdata
1+2+3     & 1660\pp72  & 3.1\pp0.3 (5.6 $\sigma$) &42\pp13  & 1149\pp4\\
1         & 1570\pp18  & 2.2\pp0.3 (3.3 $\sigma$) & 7\pp5   & 1161\pp1\\
2         & 1652\pp30  & 3.3\pp0.4 (4.8 $\sigma$) & 34\pp12 & 1144\pp4\\
3         & 1758\pp24  & $<$3.1 & & \\
rest      & 2065\pp157 & $<$1.7 & & \\
\enddata
\tablenotetext{a}{All errors correspond to $\Delta\chi^2=1$. The upper
limits are for 50 Hz FWHM and correspond to the 95 \% confidence level
($\Delta\chi^2=2.71$).}
\tablenotetext{b}{The sets used are shown in Fig.\,\ref{CD_HID}b.}
\tablenotetext{c}{The errors are the standard deviation of the count
rate distribution of the selection used. Energy range 2--18.9 keV}
\tablenotetext{d}{Energy range 2--18.2 keV.}
\end{deluxetable}

\end{document}